\documentclass[aps,prl,preprint,groupedaddress,showpacs]{revtex4}
\usepackage{graphicx}
\begin{document}

\title{Once more about  the $K\bar K$ molecule approach to the light scalars}
\author {
N.N. Achasov} \email{achasov@math.nsc.ru}
\author {
A.V. Kiselev} \email{kiselev@math.nsc.ru}

\affiliation{
   Laboratory of Theoretical Physics,
 Sobolev Institute for Mathematics, Novosibirsk, 630090, Russia}

\date{\today}

\begin{abstract}
We show that the recent paper \cite{kalash}, claiming that the
radiative decays $\phi\to a_0(980)\gamma$ and $\phi\to
f_0(980)\gamma$ should be of the same order of magnitude
regardless of whether the $a_0(980)$ and $f_0(980)$ are compact
four-quark states or  extended $K\bar K$ molecule states,  is
misleading.

\end{abstract}


\maketitle

Recently Ref. \cite{kalash} has claimed that the radiative decays
of the $\phi$ meson to the scalar $a_0(980)$ and $f_0(980)$
"should be of the same order of magnitude for a molecular state
and for a compact state...". We show below that this claim is
misleading. The authors of Ref. \cite{kalash} think that their
amplitude of the $\phi\to K^+K^-\to\gamma S$ transition (where
$S=a_0\,\mbox{or}\, f_0$) is caused by the nonrelativistic kaons
in the  $K\bar K$ molecule $S$. However, we will show below that
this is incorrect.

 Eq. (14) in
Ref. \cite{kalash}, describing the $\phi\to\gamma S$ amplitude, is
\begin{equation}
\label{yulia}
 J_{ik}=2J^{(a)}_{ik}+J^{(c)}_{ik}+J^{(d)}_{ik}=-\delta_{ik}\frac{i}{4\pi^2}(a-b)I(a\,,b\,;\Gamma)+...\,,
\end{equation}
where the subscripts $ik$  are
    the spatial Lorentz indices referring to the $\phi$ and the
    photon, $2J^{(a)}_{ik}$ corresponds to the sum of the diagrams of
Fig. 1(a) and 1(b), $J^{(c)}_{ik}$  corresponds to the diagram of
Fig. 1(c), and $J^{(d)}_{ik}$   corresponds to the diagram of Fig.
1(d) ( which is added because  "gauge invariance calls for a
correction term induced by this additional flow of charge"
\cite{kalash}  in an extended molecule case), $a=m^2_\phi/m_K^2$,
$b=m^2_S/m_K^2$. "Terms that do not contribute to the process of
interest are not shown explicitly"  \cite{kalash}. Note that Fig.
1 of our paper corresponds to Fig. 1 of Ref. \cite{kalash}.

\begin{figure}[h]
\begin{center}
\begin{tabular}{cccc}
\includegraphics[width=3.5cm]{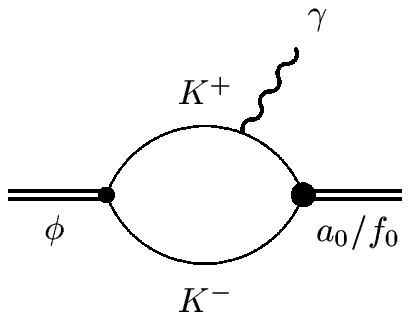}& \raisebox{-6mm}{$\includegraphics[width=3.5cm]{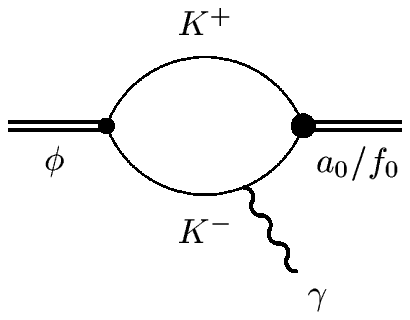}$}&
\includegraphics[width=3.5cm]{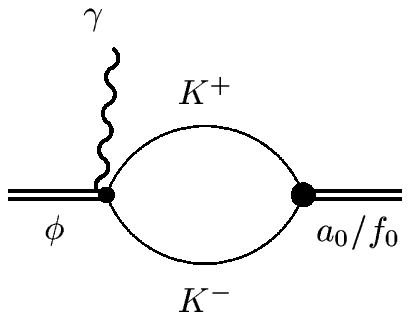}& \includegraphics[width=3.5cm]{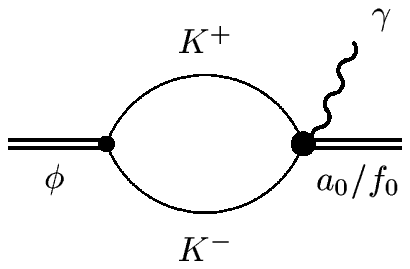}\\ (a)&(b)&(c)&(d)
\end{tabular}
\end{center}
\caption{Diagrams contributing to the radiative decay amplitude
(\ref{yulia}).}
\end{figure}

Assuming the  nonrelativistic kinematics of  kaons in the loop,
the authors of Ref. \cite{kalash} obtain  the individual integrals
\begin{eqnarray}
\label{itegrals}
 &&
2J^{(a)}_{ik}=-\frac{i}{m^3}\int\frac{d^3k}{(2\pi)^3}\frac{k_ik_j\Gamma(|\mbox{{\bf
k-q}/2} |)}{[E_V-\frac{k^2}{m}+i 0][E_S-\frac{\left
(\mbox{{\bf k-q}/2}\right )^2}{m}+i 0]}\,,\nonumber\\[3pt]
 &&
J^{(c)}_{ik}=-\frac{i}{2m^2}\delta_{ik}\int\frac{d^3k}{(2\pi)^3}\frac{\Gamma(k)
}{E_S-\frac{k^2}{m}+i 0}\,,\nonumber\\[3pt]
 &&
J^{(d)}_{ik}=-\frac{i}{2m^2}\int\frac{d^3k}{(2\pi)^3}\frac{k_ik_j}
{E_V-\frac{k^2}{m}+i
0}\,\frac{1}{k}\,\frac{\partial\Gamma(k)}{\partial k}\,,
\end{eqnarray}
where $E_V=m_V-2m>0$, $E_S=m_S-2m<0$, $m=m_K$, $V=\phi$,  {\bf q}
is a photon momentum, {\bf k} is a kaon momentum in a molecule,
$k=|\mbox{\bf k}|$, $\Gamma (\mbox{\bf k})=\beta^2/(\mbox{\bf
k}^2+\beta^2)$, $1/\beta$ is a potential range \cite{wf}. The
range of $\beta$ is typically $m_\rho\approx 0.8$ GeV, because the
$\rho$-meson exchange in the $t$-channel "is responsible for the
formation of scalars" \cite{kalash}.
 The authors of Ref.  \cite{kalash} calculate $(a-b)I(a,b;\,\Gamma)$ at {\bf q}=0
\begin{eqnarray}
\label{q=0} &&(a-b)I(a,b;\,\Gamma)=2(a-b)I^{(a)}(a,b;\,\Gamma)
+(a-b)I^{(c)}(a,b;\,\Gamma) +(a-b)I^{(d)}(a,b;\,\Gamma)\,,
\end{eqnarray}
where
\begin{eqnarray}
\label{2a,c}
 && 2(a-b)I^{(a)}(a,b;\,\Gamma)=i\, 4\pi^2\,\frac{1}{3}2J^{(a)}_{ii}
=\int\frac{1}{3m^3}\frac{k^2\Gamma(k)}{[E_V-\frac{k^2}{m}+ i
0][E_S-\frac{k^2}{m}+ i 0]}\,\frac{d^3k}{2\pi}\,,\nonumber\\[3pt]
 &&
(a-b)I^{(c)}(a,b;\,\Gamma)=i\, 4\pi^2\,\frac{1}{3}J^{(c)}_{ii}=
\int\frac{1}{2m^2}\frac{\Gamma(k) }{E_S-\frac{k^2}{m}+ i
0}\,\frac{d^3k}{2\pi}\,,
\end{eqnarray}
and
\begin{equation}
\label{d}
 (a-b)I^{(d)}(a,b;\,\Gamma)=i\,
4\pi^2\,\frac{1}{3}\,J^{(d)}_{ii}= \int\frac{1}{6m^2}\frac{k^2}
{E_V-\frac{k^2}{m}+ i
0}\,\frac{1}{k}\,\frac{\partial\Gamma(k)}{\partial
k}\,\frac{d^3k}{2\pi}\,.
\end{equation}

To reveal what kaon momenta are essential in the real part of the
$\phi\to K^+K^-\to\gamma S$ amplitude, we introduce a cutoff $k_0$
in Eqs. (\ref{2a,c}) and (\ref{d}) ($|\mbox{\bf k}|=k\leq k_0$)
and calculate the auxiliary integral
\begin{eqnarray}
\label{integralk0}
   (a-b)\mbox{Re}I(a,b;\,\Gamma\,;\,k_0)= && 2(a-b)\mbox{Re}I^{(a)}(a,b;\,\Gamma\,;\,k_0)+(a-b)\mbox{Re}I^{(c)}(a,b;\,\Gamma\,;\,k_0)\nonumber\\
&& +(a-b)\mbox{Re}I^{(d)}(a,b;\,\Gamma\,;\,k_0)
\end{eqnarray}
 for $\beta$= 0.2 GeV, 0.3 GeV and 0.8 GeV \cite{kalash,averagek2}. When
$k_0\to\infty$ the integral $I(a,b;\,\Gamma\,;\,k_0)\to
I(a,b;\,\Gamma\,;\,\infty)\equiv I(a,b;\,\Gamma)$. We use
 $m_S=980$ MeV, $m_K=495$ MeV  \cite{kalash}.
 In Fig. 2   is depicted the $\mbox{Re}I(a,b;\,\Gamma\,;\,k_0)$
dependence on a cutoff $k_0$ for different $\beta$.

\begin{figure}[h]
\begin{center}
\begin{tabular}{cc}
\includegraphics[width=7.8cm]{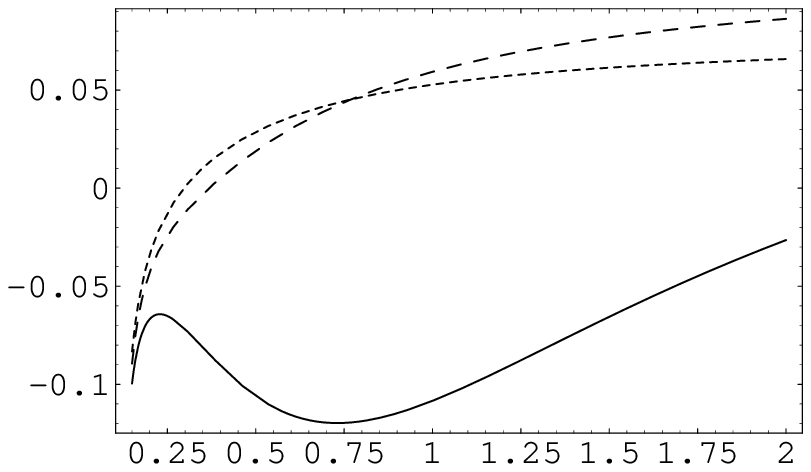}\put(-40,-10){\scalebox{1.0}{$k_0$, GeV}}&\hspace{0.5cm} {$\includegraphics[width=7.8cm]{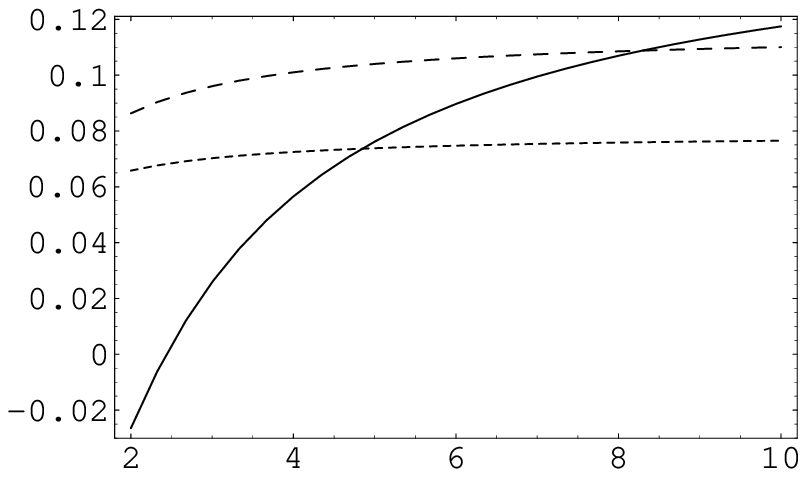}
\put(-40,-10){\scalebox{1.0}{$k_0$, GeV}}$}\\ (a)&(b)
\end{tabular}
\end{center}
\caption{$(a-b)\mbox{Re}I(a,b;\,\Gamma\,;\,k_0)$ (the definition
in the text, see Eq. (\ref{integralk0})).  The solid line for
$\beta=0.8$ GeV, the dashed line for $\beta=0.3$ GeV, the dotted
line for $\beta=0.2$ GeV. (a) $k_0\leq 2$ GeV, (b) $k_0\geq 2$
GeV. The limit values of
$(a-b)\mbox{Re}I(a,b;\,\Gamma\,;\,\infty)\equiv
(a-b)\mbox{Re}I(a,b;\,\Gamma)$ are: 0.16 for $\beta=0.8$ GeV,
0.116 for $\beta=0.3$ GeV,  0.079 for $\beta=0.2$ GeV. }
\end{figure}

 As is obvious from  Fig. 2, the contribution of
the nonrelativistic kaons ($k_0<0.3$ GeV) into
$\mbox{Re}I(a,b;\,\Gamma)$ is small in all instances. What's more
the ultrarelativistic kaons ($k_0>2$ GeV) determine the real part
of the $\phi\to K^+K^-\to\gamma S$ amplitude in the  typical case
of $\beta=0.8$ GeV, see Fig. 2(b) \cite{im}. So the authors of
Ref. \cite{kalash} use a nonrelativistic description  beyond its
region of applicability \cite{eclecticism}.

 The authors of Ref. \cite{kalash} in fact
 evaluate  the $(d)$-contribution by integrating Eq.
(\ref{d}) by parts. This gives a contribution
$(a-b)\mbox{Re}\tilde{I}^{(d)}(a,b;\,\Gamma\,;\,k_0)$ which when
summed with the $2(a-b)\mbox{Re}I^{(a)}(a,b;\,\Gamma\,;\,k_0)$ and
$(a-b)\mbox{Re}I^{(c)}(a,b;\,\Gamma\,;\,k_0)$ contributions gives
$(a-b)\mbox{Re}\tilde{I}(a,b;\,\Gamma\,;\,k_0)$ shown in Fig. 3.

 \begin{figure}[h]
\begin{center}
\begin{tabular}{cc}
\includegraphics[width=7.8cm]{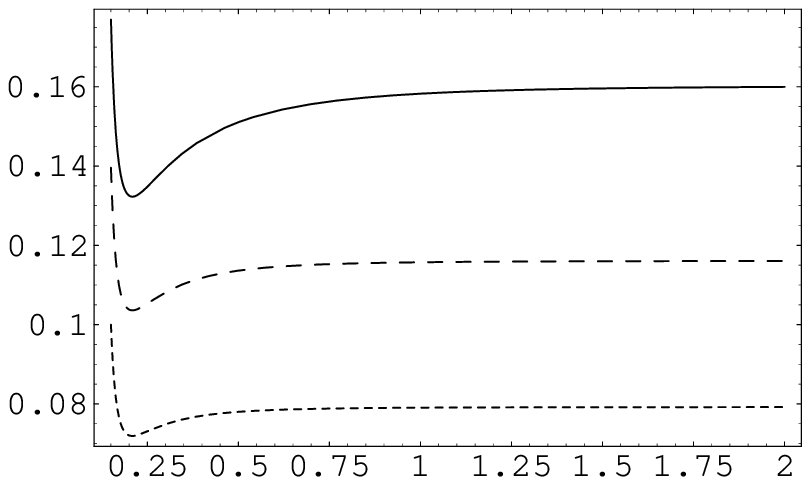}\put(-40,-10){\scalebox{1.0}{$k_0$, GeV}}&\hspace{0.5cm}
\end{tabular}
\end{center}
\caption{ $(a-b)\mbox{Re}\tilde{I}(a,b;\,\Gamma\,;\,k_0)$,
$k_0\leq 2$ GeV, $\tilde{I}(a,b;\,\Gamma\,;\,\infty)\equiv
I(a,b;\,\Gamma\,;\,\infty)\equiv I(a,b;\,\Gamma)$.  The solid line
for $\beta=0.8$ GeV, the dashed line for $\beta=0.3$ GeV, the
dotted line for $\beta=0.2$ GeV. }
\end{figure}
As is seen from Fig. 3,  the integral
$(a-b)\mbox{Re}\tilde{I}(a,b;\,\Gamma\,;\,k_0)$ converges in the
nonrelativistic region  ($k_0<0.3$ GeV). The  authors  call this
operation "a trick" \cite{kalash}  believing that the rapid
convergence of $(a-b)\mbox{Re}\tilde{I}(a,b;\,\Gamma\,;\,k_0)$
justifies their nonrelativistic approximation. But only
$(a-b)\mbox{Re}I(a,b;\,\Gamma\,;\,k_0)$ represents the momentum
 (or space) distribution of kaons and, in particular, the distribution of the charge flow  in the $K\bar
 K$-molecule and shows that the decays occur at small distances for the annihilation of
 the ultrarelativistic kaons and antikaons  ($k_0>2$ GeV) in the typical case, see Fig. 2(b).
 The difference between $(a-b)\mbox{Re}I(a,b;\,\Gamma\,;\,k_0)$
and $(a-b)\mbox{Re}\tilde{I}(a,b;\,\Gamma\,;\,k_0)$ equals the
slow convergent integral of the total derivative,
\begin{equation}
\label{td} (a-b)\mbox{Re}I(a,b;\,\Gamma\,;\,k_0) -
(a-b)\mbox{Re}\tilde{I}(a,b;\,\Gamma\,;\,k_0) = \frac{1}{3m^2}
 \int_0^{k_0}d\Biggl (\frac{k^3\Gamma(k)}
{E_V-\frac{k^2}{m}+ i 0}\Biggr )\,,
\end{equation}
which vanishes at $k_0\to\infty$. As for the finite $k_0$,
discarding  this contribution leads to a loss of physical
significance.

So, the real part of the $K^+K^-$ loop is caused by the kaon high
virtualities, that is, by  a compact four-quark system, which
points to four-quark nature of the $a_0(980)$ and $f_0(980)$
mesons \cite{achasov20012003}.

We thank C. Hanhart and Yu. S. Kalashnikova  for instructive
discussions.

This work was supported in part by  the Presidential Grant No.
NSh-5362.2006.2 for  Leading Scientific Schools and by the RFFI
Grant No. 07-02-00093 from Russian Foundation for Basic Research.
A.V.K. thanks very much the Dynasty Foundation and ICFPM for
support, too.


\begin{thebibliography}{99}
\bibitem{averagek2}
As it follows from Ref. \cite{wf} this set of $\beta$ corresponds
to the set of the nonrelativistic average momentum square of a $K$
meson in the molecule, $ < k^2 >/m_K^2$: 0.056, 0.084, and 0.224.
\end{thebibliography}
\end{document}